\documentclass[11pt]{article}
\usepackage{latexsym}
\usepackage{a4,fullpage}
\usepackage{graphicx}
\usepackage{times}
\usepackage{amssymb,amsmath}
\addtocounter{tocdepth}{3}
\title{File Transfer Application For Sharing Femto Access}
\author{Mariem Krichen, Johanne Cohen and Dominique Barth   \\ 
 PRiSM, University of Versailles, 45 avenue des Etats-Unis, 78035 Versailles, France\\
email : \{mariem.krichen\}\{johanne.cohen\}\{dominique.barth\}@prism.uvsq.fr}

\newtheorem{theorem}{theorem}
  
\date{}

\begin{document}
\maketitle


\begin{abstract}
In wireless access network optimization, today's main challenges reside in traffic offload and in the improvement of both capacity and coverage networks. The operators are interested in solving their localized coverage and capacity problems in areas where the macro network signal is not able to serve the demand for mobile data. Thus, the major issue for operators is to find the best solution at reasonable expanses. The femto cell seems to be the answer to this problematic. In this work\footnote{This work is supported by the COMET project AWARE. http://www.ftw.at/news/project-start-for-aware-ftw}, we  focus on the problem of sharing femto access between a same mobile operator's customers.  This problem can be modeled as a game where service requesters customers (SRCs) and service providers customers (SPCs) are the players.

This work addresses the sharing femto access problem considering only one
SPC using game theory tools. We consider that SRCs are static and have
some similar and regular connection behavior. We also note that the
SPC and each SRC have a software embedded respectively on its femto
access, user equipment (UE).

After each connection requested by a SRC, its software will learn the
strategy increasing its gain knowing that no information about the
other SRCs strategies is given. The following article presents a
distributed learning algorithm with incomplete information running in
SRCs software. We will then answer the following questions for a game
with $N$ SRCs and one SPC:  how many connections are necessary for each
SRC in order to learn the strategy maximizing its gain? Does this
algorithm converge to a stable state? If yes, does this state a Nash
Equilibrium and is there any way to optimize the learning process
duration time triggered by SRCs software? \\

{\bf Keywords-component: game theory,  sharing femto access,  Nash Equilibrium,  distributed learning algorithm, stable state.}
\end{abstract}

\section{Introduction}\label{sectintro}
Today, one of the biggest issues for Mobile Operators is to provide acceptable indoor coverage for wireless networks. Among the several in-building solutions,
 the femto cell  is the one which is gaining significant interest. 
A \textit{femto cell}  is a small cellular base station characterized by low transmission power, limited access to a closed user group designed for residential or small business
use. But its expansive buying cost is not motivating to purchase it. This solution could help operators solve localized coverage problems and extend their network. Indeed, in some areas where the macro network signal is weak, a network of open femto cells access would significantly improve the voice quality and data connectivity.
This would be feasible if access points owners accept to be part of a Club where each member is willing to open up its access point to the other members. This idea of sharing part of its bandwidth started with FON\footnote{FON is a for-profit company incorporated and registered in the U.K. FON was created in Madrid, Spain, by Martin
  Varsavsky, an Argentine/Spanish entrepreneur and founder of many
  companies in the last 20 years.}, a club where members share their WiFi connection and inspired us to propose a Club where members share their femto accesses with bandwidth guarantees. A femto club member could share its 3G/LTE signal securely with other club members. Incentives for an owner of an access point to be member of such a club can be not only to share a part of the cost of the access point but also to make advertisement or to share some information through a specific social network associated to the club. These incentives would logically lead the club to manage by itself only such bandwidth exchange, but since this technology uses licensed spectrum, the only model that could for the moment be adopted for sharing femto access is the one where the mobile operator is also participating. Sharing femto access is a service proposed by the Mobile Operator to its clients. These customers are divided into  \textit{ service providers customers (SPCs)} and  \textit{service requesters customers (SRCs)} : \textit{SPCs} are the owners of femto cells accesses for which they have contracted with a\textit{ Mobile operator} denoted by $MO$. \textit{SRCs} are customers using a mobile terminal in an area covered by some SPCs access points and requesting to use these access points. 
Note that a user can be both a SPC and a SRC. Dynamic femto spectrum sharing is a challenging problem for all the actors. 
Indeed the amount of requested bandwidth by SRCs as well as the amount of bandwidth shared by SPCs and pricing should be determined such that the utility of all 
the agents is maximized. Since the interests of all the actors could be antagonist, especially between many SRCs requesting a same SPC, we model our system as a game to determine equilibria of such situations. 

\paragraph{\bf Related works}
\leavevmode\

\label{sec:relatedWorks}
The problem of sharing bandwidth and pricing has been already addressed by Dusit Niyato et al \cite{papa}, then modeled as a game. The challenging problem in this context is that bandwidth sharing requires a “peaceful” co-existence of both primary and secondary users.
The femto access sharing we present in this article is similar but takes also into account both of SRCs and SPCs profiles.  

The  potential games introduced by Rosenthal~\cite{rosenthal73}   are classical games  having at least  one pure Nash equilibrium.  These games have a potential function such that each of its local optimums corresponds to a pure Nash equilibrium. This property has been used 
 for congestion game in general (see \cite{AlgoGameTheory} for a survey), with Resource Reuse in a wireless context (see \cite{citation:6}) and for a real-time spectrum sharing problem with QoS provisioning~\cite{citation:7}.

A decentralized learning algorithm of Nash equilibria in multi-person stochastic games with incomplete information has been presented by M.A.L. Thathachar et all. In the considered game, the distribution of the random payoff is unknown to the players and further none of the players know the strategies or the actual moves of other players. It is proved that all stable stationary points of the algorithm are Nash equilibrium for the game \cite{citation:3}. The study presented in this article will use this algorithm in the game restricted to SRCs where each SRC will learn the strategy maximizing its gain using only local information. We will check whether if the stationary point the algorithm converges to is a pure Nash equilibrium.

\paragraph{\bf Our contribution.}
\label{sec:Contribution}
 Section~\ref{sec:gener-model-shar} presents the model for sharing femto access and the actors involved in this model are described.
The game considering only SRCs competition is presented in Section~\ref{sec:game-presentation}. Section~\ref{sec:learning-level-1}
details the principle of a distributed algorithm used to learn the game NE if any exists and some simulations results are given in Section~\ref{sec:conclusion Simulations}. Finally, Section~\ref{sec:perspectives} draws a general conclusion and gives some perspectives.

\section{Model for sharing femto access}
\label{sec:gener-model-shar}

Now, we present the model for sharing femto access.  Our system is
composed of SPCs and SRCs. SPCs share some amount of bandwidth with
SRCs. We assume that there exists a Token Based Accounting Protocol to
exchange services between SRCs and SPCs against
tokens~\cite{citation:1}.  On one hand, the paradigm for spectrum
sharing should guarantee a certain access speed and a data
transmission speed. On the other hand, the model should propose a type
of connection (that would be more expansive than others) which
guarantees QoS to SRCs: connections belonging to this type will never be canceled by the
SPC.

Our work considers a unique SPC denoted by $X$ and several SRCs. We
assume that the SPC's resource reserved for sharing is an amount of bandwidth
denoted by $B_{S}(X)$. Then, we will consider that each SRC $Y$ requests connection
from $X$. Actually, the bandwidth $B_{S}(X)$ of SPC $X$ is divided into two
parts:

      $$B_{S}(X)=B_{S_{G}}(X)+ B_{S_{Y}}(X) $$

\begin{itemize}
\item $B_{S_{G}}(X)$ is the part of bandwidth in which SRCs communications  can never be preempted. 
 It is kind of a guaranteed QoS allocated to SRCs.
\item$ B_{S_{Y}}(X)$ is the part of bandwidth in which  SRCs communicatiosn  can  be preempted. 
This preemption is due to the fact that the SPC has priority on this part of bandwidth.
Thus, a communication allocated in $B_{S_{Y}}(X)$ is characterized by a risk of preemption.
\end{itemize}

\begin{figure}[Ht]
\begin{center}
\includegraphics[height=5cm,width=7cm]{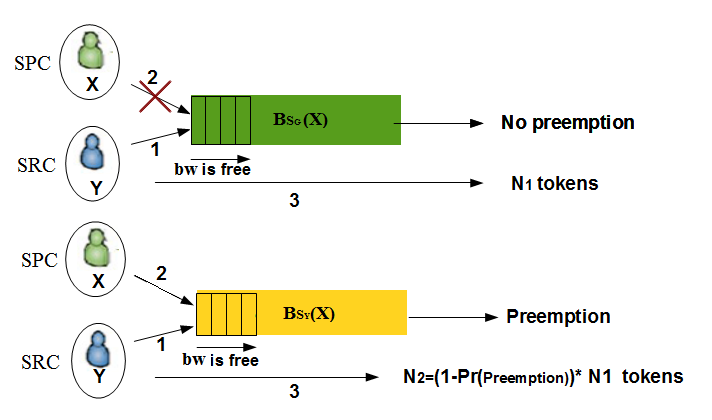}
\end{center}
\caption{Bandwidth sharing process and billing process}
\label{fig:1}
\end{figure}

Figure \ref{fig:1} introduces the sharing bandwidth process and the
billing process in both cases of a \textit{Green} and a
\textit{Yellow} connection allocation: let's consider a SRC $Y$ who
needs an amount of bandwidth equal to $bw$. $bw$ will be allocated to $
Y$ only if it is free. If $X$ will need $bw$, he will not be able to
use it if the connection he allocated to $Y$ is green. However, he
will be able to preempt the connection allocated to $Y$ if it is a
yellow one.

We assume that there exists a Token Based Accounting Protocol to exchange services between SRCs and SPCs against
tokens (representing money)~\cite{citation:1}. The billing process depends on whether  
the SRC has used a green connection or a yellow one.

In the case of a Green connection, the SRC will spend $N_1$ tokens
corresponding to the used bandwidth $bw$.  In the case of a Yellow
connection, the SRC will spend $N_2$ ($N_2< N_1$) tokens if the
connection succeed. Indeed, the yellow connection is cheaper than the
green one due to the risk of  preemption. If the yellow connection
given to the SRC has failed, this connection will be free.

\label{sec:adist-algorithm}

\subsection{Actors Description}\label{sec:actor-description}

In the following section, we will describe the actors involved in the model that we have just presented.

\subsubsection{SPC Actor} \label{sec:spc-actor}
 \leavevmode\

A SPC is a customer of the MO. SPC proposes to share an amount of its bandwidth
with SRCs for a price per bandwidth unit which depends on the type of
connection (Green,Yellow). The bandwidth split into Green and Yellow
parts is determined following its \textit{sensitivity to Gain} denoted by $ \mu
\in[0,1]$ and its \textit{sensitivity to its own connection QoS} denoted by
$\Gamma \in[0,1]$. These two parameters are dual: $\mu+ \Gamma =1$.

The \textit{Gain sensitivity} parameter indicates its sensitivity degree to the
price of the connection shared while the \textit{QoS sensitivity} parameter
indicates the SPC's tolerance degree towards preemption risk. The more
a SPC is sensitive to gain, the bigger its green part bandwidth will
be because of its expansive selling price. The more a SPC is
sensitive to QoS, the bigger its Yellow Part bandwidth will be since
he does not want to be preempted by SRCs. When $\mu>1/2$, we say that the
SPC is sensitive to gain. Otherwise, the SPC is considered as
sensitive to its access QoS.

\subsubsection{SRC Actor}\label{sec:src-actor}
\leavevmode\

A SRC is also a customer of the MO in need of QoS characterized by a\textit{ QoS sensitivity} parameter $\alpha$ and a\textit{ price sensitivity} parameter $\beta$. The SRC wants to use the SPC's resources.
 The \textit{QoS sensitivity} parameter indicates the SRC's
tolerance degree towards the QoS degradation while the \textit{price
sensitivity} parameter indicates the SRC's tolerance degree towards the cost of
the connection. 
These two parameters are dual:
$\alpha+ \beta=1$.

The more one SRC is sensitive to QoS, the less its tolerance degree
towards the QoS degradation will be. The more one SRC is sensitive
to price, the less he is able to pay for the connection. When
$\alpha>1/2$, the SRC is considered as sensitive to QoS. Otherwise,
the SRC is sensitive to price.

\subsubsection{Interaction between the SPC actor and SRCs actors}\leavevmode\

\label{sec:interaction}
In general context, SRCs will request for some amount of bandwidth from the SPC depending of their profiles. The SPC will treat the SRCs requests for a fixed bandwidth split.
\begin{enumerate}
\item For SRCs, the utility depends on their requests, the other SRCs requests as well as the SPC's decision (bandwidth allocated and type of connection) for a fixed SPC's bandwidth split. So, the SRC's utility depends mainly on its competition with other SRCs to receive some bandwidth from the same SPC.
\item For the SPC, the utility depends on its bandwidth split for fixed SRCs requests. Since we consider only one SPC, no competition is needed for the SPC to rise its utility.
\end{enumerate}
 
In our work, we will only focus on the competition between SRCs when the SPC's bandwidth split is fixed.

\section{Game presentation}
\label{sec:game-presentation}

Since only the competition between SRCs is considered, sharing femto access problem could be thus modeled as a game where $N$ SRCs
are the players. We also consider that the SPC's bandwidth split is fixed. This could be motivated by the fact that learning methods are reliable only when the surrounding environment is invariant.
We do not consider the mobility of SRCs. However, we assume that SRCs have some regular and similar connection
behavior: each SRC requests the SPC's connection in nearly same time
slots with almost invariant needs in terms of QoS. This means that
requesting the SPC's femto access becomes almost routine for SRCs. This could be seen as repeated games.

Along requested connections, each SRC
will learn, thanks to an algorithm running in a software
embedded in its user equipment, the best strategy to be played to
maximize its gain.
In this article, sharing femto access game will be denoted by the game restricted to SRCs.



The game restricted to SRCs is defined as follows: given a
fixed SPC's bandwidth split (into green part and yellow part), what
would be the best strategy to be played by SRCs in order to have a
stable situation where the strategy of each SRC player is optimal for him considering the other SRCs strategies. This situation corresponds to a pure Nash equilibrium  in game theory. 
Recall that a \textit{pure Nash equilibrium} of a game is a situation where, for each player, there is no
  unilateral strategy deviation that increases its utility \cite{citation:2}.

 

\subsection{Game restricted to SRCs}
\label{sec:level-1-game}

As mentioned in the previous section, the game restricted to SRCs assumes
that the SPC's bandwidth split is fixed. Let $B_S$ be the total
bandwidth the SPC is agree to share and let $\Psi_S$ be the proportion of
$B_{S_{G}}$ regarding $B_S$.

\subsubsection{ SRCs  QoS needs}
\leavevmode\

\label{sec:src-modelisation}

Each SRC's QoS needs depend on the type of application he requests.
Requesting for femto access is equivalent to request an amount of bandwidth. Fixing this amount of bandwidth depends on the following parameters: the type of application (real time, elastic), the QoS parameter that the applications requires (delay, time transfer file, \dots ), the type of connection (Green or Yellow) and the SRC's profile (QoS sensitive SRC or price sensitive SRC).

Our work takes into account the File Transfer Application. QoS is defined as the time transfer file that we will denote by $t$. The SRC's QoS satisfaction is related to $t$.
For each SRC, the time transfer file should be between $T_1$ and $T_2$ and is defined as follows:

\begin{itemize}
\item  Case  $t=T_1$:  $BW_{Max}$ corresponds to the required bandwidth to download a file in $t=T_1$. If the SPC provides an amount of bandwidth equal to $BW_{Max}$, then the SRC's QoS satisfaction is at the top.
\item  Case  $t=T_2$:  $BW_{Min}$ corresponds to the required bandwidth to download a file in $t=T_2$. If the SPC provides an amount of bandwidth  equal to $BW_{Min}$, then the SRC's QoS satisfaction is minimal.
\end{itemize}
Each SRC 
will request for a minimum amount of bandwidth and a maximum amount of bandwidth in Green and Yellow depending on its profile and on its user equipment Signal-Strength towards the SPC's femto cell.

All the SRCs requests can not be accepted. In fact, 
since the SPC's bandwidth is limited and since several SRCs could request for the SPC's connection at the same time, one possible response that a SRC could receive is a deny one. Requesting for a Minimum and a Maximum amount of bandwidth will decrease the chances to receive such a response. Besides, requesting a bandwidth interval generalizes the fact of requesting a fixed amount of bandwidth. In this way, a SRC could receive an amount of bandwidth which may be different from its optimal request but would avoid him to have no payoff.

The minimum and the maximum amount of bandwidth are fixed depending on
the SRC's profile. Note that the parameters caracterizing a SRC's
profile are real in $[0,1]$. We aim at translating these
parameters into intervals of bandwidth requests which are actually
integers. So, we introduce a parameter $\varepsilon$ representing
the discretization of the bandwidth resquested. Let  $S_{SRC_{i}}$be the set of possible strategies of $SRC_i$.
In the following, we focus on the request of a SRC denoted by $SRC_i$ according to its profile.

\begin{enumerate}
\item We consider the case where $SRC_i$ has its QoS sensitive parameter $\alpha_i$ greater than $1/2$.  $SRC_i$ fixes a revenue threshold  under which he denies any proposed connection (high QoS degradation). This threshold  denoted by $Rev\_Th_i$  corresponds to a minimum amount of bandwidth to be requested. This parameter depends on the QoS sensitivity  $\alpha_i$ of the SRC.  $Rev\_Th_i=\alpha_i- \kappa $ where $\kappa\in[0,1]$ is the allowed variation from the QoS degradation tolerance fixed according to the SRC's profile (more specifically $\alpha_i$).

\begin{center}
$S_{SRC_{i}}= \{Rev\_Th_i ,Rev\_Th_i+\varepsilon,Rev\_Th_i+2\varepsilon,\dots,1\}$.
\end{center}

\begin{figure}[Ht]
\begin{center}
\includegraphics[height=5cm,width=8cm]{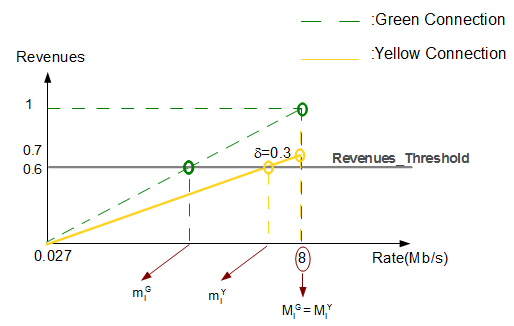}
\end{center}
\caption{Bandwidth request illustration for a QoS sensitive SRC's strategy ($\alpha=0.7$, $\varepsilon=0.1$, and $\kappa=0.1$). }
\label{fig:8}
\end{figure}

\leavevmode\

\item We consider the case where  $SRC_i$ has its QoS sensitive parameter $\alpha_i$ less than $1/2$. This implies that its price sensitive parameter $\beta_i$ is greater than $1/2$.
 $SRC_i$ fixes a cost   threshold     denoted by $Cost\_Th_i$  above which he denies any proposed connection (the cost is beyond what he is    able to pay). This threshold corresponds to a maximum amount of bandwidth to be requested. This parameter depends on the QoS sensitivity of the SRC and is defined as follows: $Cost\_Th_i= \alpha_i+\kappa $.

 \begin{center}

$S_{SRC_{i}}=\{ Cost\_Th_i ,Cost\_Th_i-\varepsilon,Cost\_Th_i-2\varepsilon,\dots,0\}$.
 \end{center}
\begin{figure}[Ht]
\begin{center}
\includegraphics[height=5cm,width=8cm]{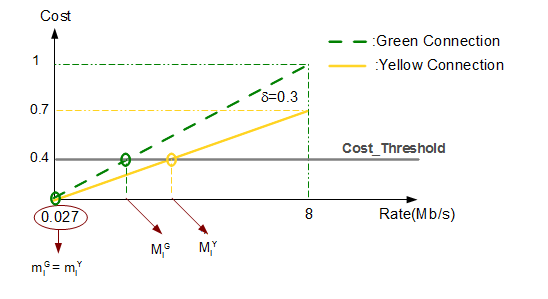}
\end{center}
\caption{Bandwidth request illustration for a price sensitive SRC's strategy ($\alpha=0.3$, $\varepsilon=0.1$, and $\kappa=0.1$).}
\label{fig:9}
\end{figure}

\end{enumerate}


Each element $s_i$ of $S_{SRC_{i}}$ permits to define an interval of
bandwidth to be requested in Green and Yellow connection. 
This represents a couple $(g_i =[m_i^G,M_i^G ],y_i =[m_i^Y,M_i^Y])$ of couples of integers. The parameters
 $m_i^X ,M_i^X$ represent respectively the minimum and the maximum amount of bandwidth to be requested in $X$ connection where $X\in \{G,Y\}$.
They are defined as follows :

\begin{enumerate}
\item Case where $SRC_i$ has its QoS sensitive parameter $\alpha_i$ greater than $1/2$:
  \begin{enumerate}
  \item   $m_i^G=\frac{BW_{max}}{s_i}$ and  $m_i^Y=\frac{BW_{max}}{s_i}\times (1-\delta)$ 
  \item   $M_i^G= BW_{max} $ and  $M_i^Y= BW_{max}$ 

  \end{enumerate}

\leavevmode\

\item Case where  $SRC_i$ has its QoS sensitive parameter $\alpha_i$ less than $1/2$.  
 \begin{enumerate}
  \item   $M_i^G=\frac{BW_{max}}{s_i}$ and  $M_i^Y=\frac{BW_{max}}{s_i}\times (1-\delta)$ 
  \item   $m_i^G= BW_{min} $ and  $m_i^Y= BW_{min}  $ 

  \end{enumerate}

\end{enumerate}

Figure \ref{fig:8} highlights an example of a QoS sensitive SRC ($\alpha=0.7$) and Figure \ref{fig:9} gives an example of a price sensitive SRC ($\alpha=0.3$).
They show the minimum and the maximum amount of bandwidth to be requested in Green and Yellow for one of the SRC's strategies. 

\subsubsection{ SPC's  bandwidth allocation}
\leavevmode\

Each SRC sends a request to the SPC. At reception, the SPC decides the way its bandwidth is allocated to SRCs. The request of each $SRC_i$
is represented by one element in $S_{SRC_{i}}$.  According to a set $\Pi$ of SRCs requets $\Pi=<s_1,s_2,\dots,s_N>$  where $s_i$ corresponds to the request of $SRC_{i}$, for any $i$, $1\leq i \leq N$, 
 SPC gives an answer  to each $SRC_i$ represented by a triple  $(G_i,Y_i,bw_i)$ defined as follows:

\begin{itemize}
  \item $bw_i$ represents the amount of bandwidth given by the SPC to $SRC_i$. 
  \item  $G_i =1$ (resp. $Y_i =1$)  means that $SRC_i$ has received a Green (resp. Yellow) connection from the SPC.  Note that the case where $G_i =1$ and $Y_i =1$ is not possible. 
  \item  If $G_i=0$ and if $Y_i=0$, then $SRC_i$ has received no
  connection from the SPC.
  \end{itemize}

Let $config(\Pi)$ be the set of all answers (one answer per SRC) of the SPC to $\Pi$. In other words,

    $$config(\Pi)=<(G_1,Y_1,bw_1),(G_2,Y_2 ,bw_2),\dots,(G_N,Y_N,bw_N)>$$  
The answers respect the two following  properties. Let  $s_i$ be the corresponding bandwith request   $(g_i =[m_i^G,M_i^G ],y_i =[m_i^Y,M_i^Y])$ of $SRC_i$. 

\begin{enumerate}
\item The SPC gives $SRC_i$ an amount of bandwidth equal to $bw_i$ where $bw_i$ is in the interval requested. More formally,  if ($G_i=1$) then $bw_i \in g_i$ or 	if ($Y_i=1$) then $bw_i \in y_i$.
\item The SPC provides bandwidth to SRCs in the limits of its bandwidth availability in Green and Yellow. Thus:
\begin{center}
 	$\sum_{i=0}^N	G_i \times  bw_i <\Psi_S B_S$ and $\sum_{i=0}^N Y_i\times   bw_i <(1-\Psi_S) B_S$.
\end{center}
\end{enumerate}

Moreover, the SPC  allocates its bandwidth in a way maximizing this outcome function:
\begin{equation}
  \label{eq:SPC}
  U_{SPC}(config(\Pi))=\sum_i Prop(bw_i) (\mu-\Gamma) \left( G_i  + Y_i  (1-\delta)  \right) 
\end{equation}
  
 Note that the SPC's outcome function considers only its own profile described in Section~\ref{sec:spc-actor}.
 Given a SRC strategy set, the SPC allocates to each SRC a connection in Green and Yellow such as its outcome function is maximized.

\subsubsection{SRC Game Definition}
\leavevmode\

\label{sec:src-game-definition}
Game theory models the interactions between players competing for a common resource.  In our system, the formulation of this noncooperative game $G=\left\langle \mathcal{N}, S, {U_k}\right\rangle$ can be described as follows:
\begin{itemize}
\item  The set of players is  $\mathcal{N}$. Each player is a SRC. There are $N$ SRCs.
\item  The space of pure strategies $S$ formed by the Cartesian product of each set of pure strategies $$S=S_{SRC_{1}}\times S_{SRC_{i}} \times ... \times S_{SRC_{N}}$$
Note that for each ${SRC_{i}}$, the set $S_{SRC_{i}}$ is defined in Section \ref{sec:src-modelisation}. A pure strategie $s_i$ is a value corresponding to the request which  is a couple $(g_i =[m_i^G,M_i^G ],y_i =[m_i^Y,M_i^Y])$ of couples of integers. 
\item  A set of utility functions $\{U_1,U_2,...,U_{N}\}$ that quantifies the players' preferences over the possible outcomes of the game.
 According to a set $\Pi$ of SRCs requets $\Pi=<s_1,s_2,\dots,s_N>$  where $s_i$ corresponds to the strategy of $SRC_{i}$, for any $i$, $1\leq i \leq N$, the SRCs utilities are determined through the $SPC$'s allocation. 
Since several allocation decisions could maximize the SPC's outcome function given by Equation \eqref{eq:SPC}, the SRC's  utility corresponds to a mean of all these allocation decisions.  Let $sol(\Pi)$ be the set of $config(\Pi)$ that maximizes the SPC's outcome function with $M=|sol(\Pi)|$.
 
The utility $U_i(\Pi)$ of $SRC_i$ from $sol(\Pi)$ is expressed as follows:

\begin{equation}
  \label{eq:SRC}
  U_i(\Pi)=\sum_{c \in sol(\Pi) }{\frac {gain_i(c)}{M}}.  
\end{equation}

  The gain of  $SRC_i$ from the SPC's allocation decision $c$ in $sol(\Pi)$  is expressed as follows:

  \begin{center}
    $gain_i(c)=Rev_i(c)-Cost_i(c)$
  \end{center}
Where $c$ represents a triple $(G_i,Y_i,bw_i)$ to $SRC_i$.
\begin{itemize}
\item 
 $Rev_i(c) \in[0,1]$ represents the SRC's QoS satisfaction from the connection $c$ provided by the SPC and is expressed as follows:
 
$$Rev_i(c)=Rev(bw_i) G_i \alpha_i+Rev(bw_i) Y_i (1-\delta) \alpha_i$$ where $Rev(bw_i) \in[0,1]$
\item  $Cost_i(c) \in[0,1]$  represents the cost of the connection $c$ provided by the SPC and is expressed as follows:

$$Cost_i(c) =Cost(bw_i) G_i \beta_i+Cost(bw_i) Y_i (1-\delta) \beta_i$$ where  $Cost(bw_i) \in[0,1]$ 
\end{itemize}
 
 We normalize the utility of $SRC_i$ in order to have $ U_i(\Pi) \in[0,1]$. 
 
\end{itemize}

\subsection{ SRC game equilibrium}
\label{sec:src-game-equilibrium}
Since each $SRC_{i}$ has a finite set of strategies, this game has a
mixed Nash equilibrium~\cite{citation:2}. In the following, we study
the existence of a pure Nash equilibrium in the sharing femto access
game using the properties of potential games. The definition of
potential game could be found in \cite{rosenthal73}.

\begin{theorem}
\label{th:TH1}
Each instance of the game restricted to SRCs admits at least one pure
Nash equilibrium.
\end{theorem}
The proof of this theorem is detailed in \cite{RR:Mariem}. 
 
\paragraph{\bf Sketch of proof:} The main idea of this proof is to
show that the Best Response Dynamic in this game converges to a pure
Nash equilibrium. The Best Response dynamic algorithm corresponds to
a sequence of profiles computation. Let $C$ be an arbitrary profile. If no
player has incentive to change its strategy in $C$, then $C$ is a pure
Nash equilibrium and this algorithm stops.  If at least one player has
incentive to change its strategy in $C$, then this player changes its
strategy by choosing its best response and thus we move to a new profile
$C'$. Then, the algorithm applies the same process for $C'$ and so on.
If some players have incentive to change their strategy, this means that the SPC
reduces its non reserved bandwidth. Thus the Best Response Dynamic
algorithm will end up since at each step the free SPC's bandwidth is
reduced.

 \section{Learning the Game restricted to SRCs' Nash equilibrium}
In Section~\ref{sec:game-presentation}, we have proved that the game restricted to SRCs admits at least one pure Nash equilibrium. Now, we want to know whether a distributed algorithm (each player knows only its local information) could converge to a pure Nash equilibrium.

\label{sec:learning-level-1}

 \label{sec:algorithm}

\subsection{Algorithm Principle}
In the following section, we will assume that we are in a distributed context. So, each SRC player will, based only on its local information, learn the strategy maximizing its gain. 
\begin{figure}[Ht]
\begin{center}
\includegraphics[height=6cm,width=7cm]{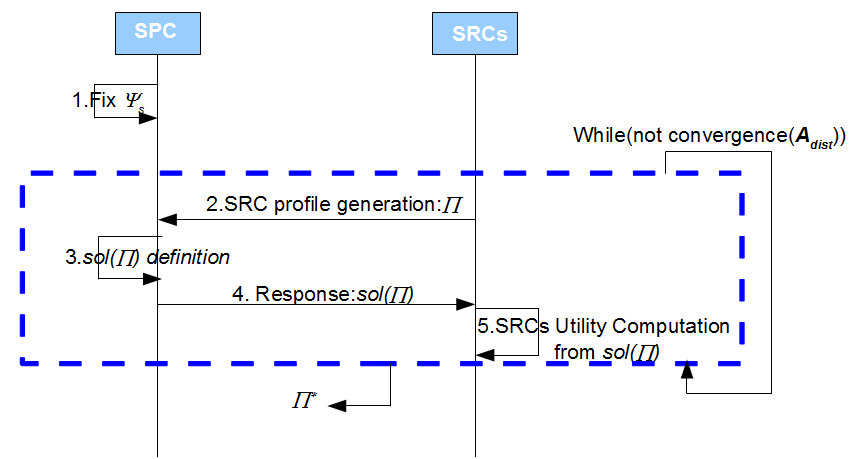}
\end{center}
\caption{Algorithm principle of the game restricted to SRCs}
\label{fig:2}
\end{figure}

For a fixed $\Psi_S$, we will apply the steps from 2 to 5 of the algorithm presented in Figure \ref{fig:2}

\begin{itemize}
\item[2.]  Each SRC sends a request using a specific distributed algorithm  denoted by $A_{dist}$.
\item[3.]  The SPC defines its decision $sol(\Pi)$. 
\item[4.] The SPC sends its decision to all SRCs.
\item[5.] Each SRC will compute its utility following $sol(\Pi)$. 
\end{itemize}
 
We will repeat all these steps till convergence of Algorithm $A_{dist}$.
We will denote by $\Pi^*$ the SRC strategy profile for which $A_{dist}$ converges. Now, we will present the Algorithm $A_{dist}$. In [3], it has been proved that if the considered game has at least
one pure Nash equilibrium and if there exists a sufficiently small value of the
learning speed parameter for which the distributed learning algorithm
converges, then the point of convergence of this algorithm is a pure Nash equilibrium. We say that this algorithm weakly converges to a Nash equilibrium.

\subsection{ALGORITHM $A_{dist}$}

Algorithm $A_{dist}$ is a decentralized learning algorithm of Nash
equilibrium in multi-person stochastic games with incomplete
information. The principle of $A_{dist}$ is described as follows:

We consider that the strategic process for each SRC player follows a discrete learning technique.
\begin{itemize}
\item Each $SRC_i$ will update its strategy at a step $t$ denoted by $s^t_i$
following its local mixed strategy (probability distribution assigned to each available pure strategy). Each $SRC_i$ will thus compute
its new probability vector $s^{t+1}_i$ using only a set of local information ($s^t_i,\ margin^t_i,\ gain^t_i$) where $gain^t_i$ represents the gain from the SPC's allocation decision and $margin^t_i$ is the pure strategy in $S_{SRC_i}$

\item At each step $t$, each SRC chooses randomly one strategy ($margin^t_i$) and increases its probability of choosing this strategy($s^t_i$) according to its gain ($gain^t_i$) and a learning parameter $b \in[0,1]$ which modulates the learning speed of the different SRCs players.

The learning technique is based on the following update rule:

\begin{center}
  \begin{tabular}[h]{lll}
    If $j \neq margin^t_i$ & then & $s^{t+1}_{i,j}=s^t_{i,j} -b.u^t_i.s^t_{i,j}$ \\
 & else &   $s^{t+1}_{i,j}=s^t_{i,j} -b.u^t_i.\sum_{k \neq margin^t_i} s^t_{i,k}$ \\
  \end{tabular}
  
\end{center}

Such that:
$u^t_i=\frac{ gain^t_i-A^t_i}{U^t_i-A^t_i}$ is the normalized utility.

The variables $(U^t_i = max_{k\leq t}  gain^k_i )$ and $(A^t_i = max_{k\leq t}  gain^k_i)$ correspond respectively to the maximum utility and the minimum utility of $SRC_i$ at iteration t. Note that it is possible to consider $A^t_i =0$

\end{itemize}

The aim from using this update rule is the following: each SRC player lowers the probabilities associated to margins not played at the step t by the same percentage. The sum of these percentages is then added to the probability associated to the margin played at step $t$.

As mentioned in  Theorem \ref{th:TH1}, in the game restricted to SRCs, there is at least one pure Nash equilibrium. In this article, we will try to answer the following questions: Does $A_{dist}$ converge? Is the point of convergence if any exists a pure Nash equilibrium? Is this algorithm reliable for Nash equilibrium computation?

\subsection{Simulations}
\label{sec:simulations}

In our simulations we will consider $N$ SRCs: each $SRC_i$ is characterized by $\alpha_i$ (the QoS sensitivity parameter of $SRC_i$) and is requesting for the SPC's connection characterized by $\mu$ to download a file.

We will only focus on the cases of a same category SRCs and more specifically QoS sensitive SRCs. Indeed, this case is the hardest one to reach stability since the SRCs requirements are conflicting. In the other possible cases, Algorithm $A_{dist}$ converges.

The simulations presented take into consideration an extremely gain sensitive SPC (i.e $\mu=1$) and $N$ QoS sensitive SRCs. The $N$ SRCs want to download a file of $1 Mbyte$. We will consider $T_2=300sec$, $\mu=1$, $\delta=0.1$, $\varepsilon=0.1$, $\kappa=0.1$, $\Psi_S =0.5$ and $B_S=20Mb/s$. Since a femto access could support only $8$ communications, we will run simulations where $2\leq N\leq 8$ keeping the same input presented above. In this article, we will present the results only for $4$ and $5$ SRCs. The first simulation presents a scenario where SRCs have only two strategies (extremely QoS sensitive SRCs\footnote{An extremely QoS sensitive SRC is a SRC with $\alpha=1$}). In the second simulation, SRCs have more than two possible strategies. In our simulations, we consider the following strategy notation $s_i=Rev\_Th_i$

   \subsubsection{Scenario 1 with 5 SRCs}
  \label{sec:sce-1-5}
\leavevmode\

\begin{figure}[Ht]
\begin{center}
\includegraphics[height=5cm,width=10cm]{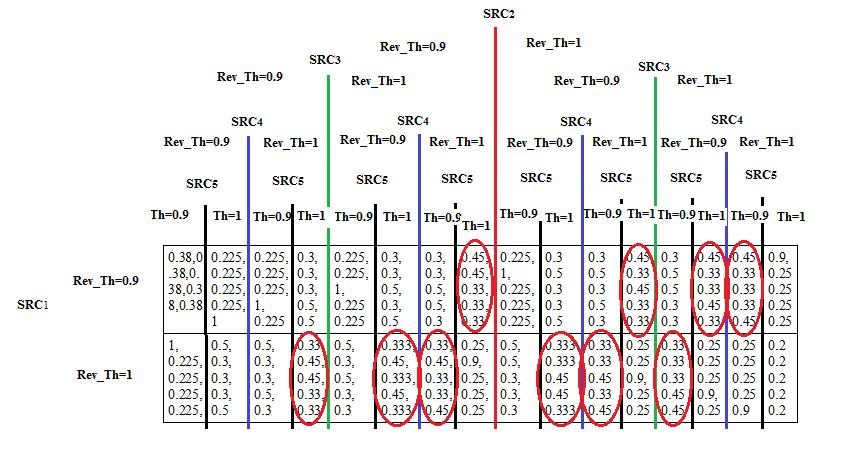}

\end{center}
\caption{SRCs Utility Matrix for Scenario 1}
\label{fig:5}
\end{figure}

\begin{figure}[Ht]
\begin{center}
\includegraphics[height=5cm,width=10cm]{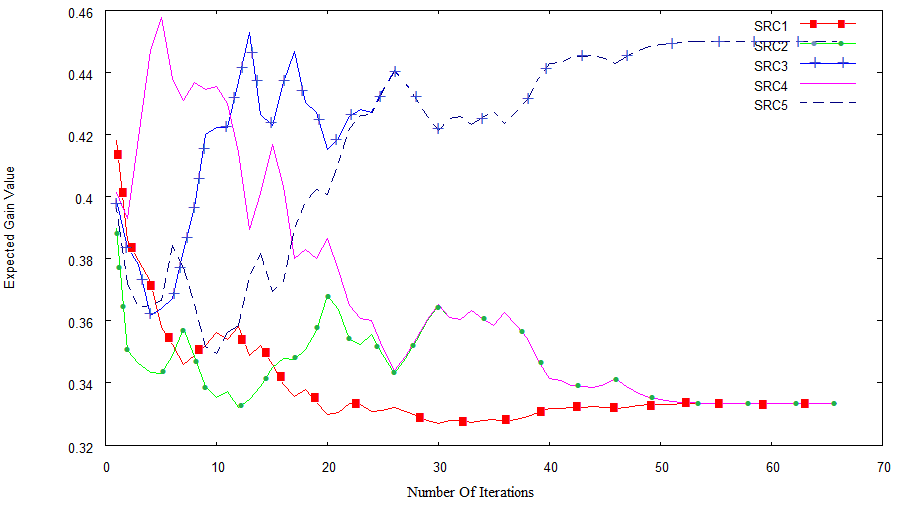}

\end{center}
\caption{Variation of SRCs expected gain in scenario 1}
\label{fig:3}
\end{figure}

  For this scenario, we will as a first step analytically compute the  game Nash equilibriums. To do so, we will consider the same steps presented in Figure \ref{fig:2} except that the SRC strategy profile is not generated with Algorithm $A_{dist}$: we will consider all the possible SRC strategy profiles and thus fill the SRCs payoff matrix with utilities corresponding to each SRC strategy profile. Figure \ref{fig:5} highlights 10 pure Nash equilibriums circled in red. These Nash equilibriums are detected analytically through the SRCs payoff matrix since SRCs cannot rise utility by an unilateral deviation.
     Now, we will run the Algorithm $A_{dist}$ . First, we will verify whether if it converges and then we will check
  whether if the point of convergence ,if any, is one among the pure Nash equilibriums
  detected analytically. To do so, we will consider $b=0.1$ and $1000$
  iterations. In Figure \ref{fig:3} and \ref{fig:4}, each point represents the mean over 15 iterations of respectively SRCs expected gain value and SRCs strategy probability.  
 
\begin{figure}[Ht]
\begin{center}
\includegraphics[height=5cm,width=10cm]{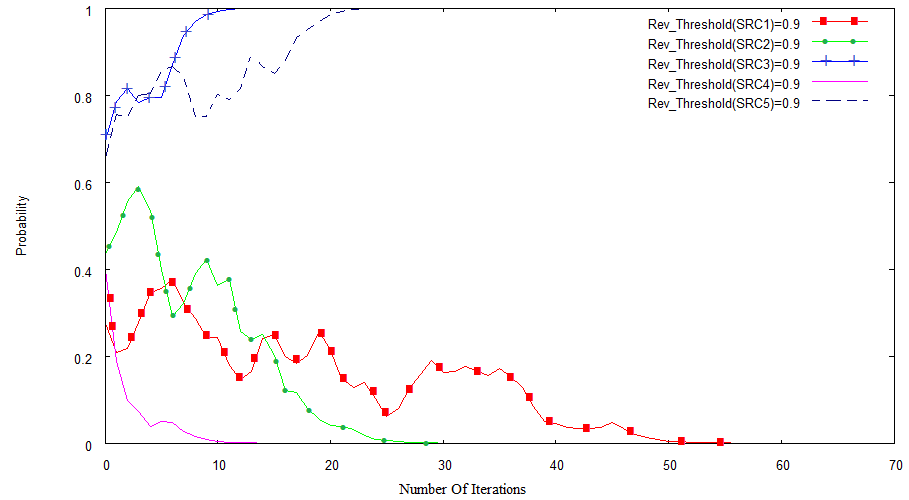}

\end{center}
\caption{Variation of SRCs strategies probabilities for Scenario 1}
\label{fig:4}
\end{figure}
  
$A_{dist}$ converges after $780$ Iterations. Figure \ref{fig:3} shows that the SRCs expected gain stabilize as follows:

Expected gain(SRC$_{1}$)=Expected gain(SRC$_{2}$)=Expected gain(SRC$_{4}$)=0.45 and Expected gain(SRC$_{3}$)=Expected gain(SRC$_{5}$)=0.33.

In Figure \ref{fig:4}, we remark that the convergence point (point where each SRC has a pure strategy) $\Pi^*=(1,1,0.9,1,0.9)$ matches with one of the pure Nash equilibriums computed analytically.

For a number of SRCs varying from $2$ to $8$, we have checked by simulations that pure Nash equilibrium in the game restricted to SRCs is reachable keeping the same input as scenario 1. 

The following table summarizes the number of iterations necessary for $A_{dist}$ to converge for $N$ varying from $2$ to $8$. This result is true for SRCs with only two strategies.

\begin{table}[h]
  \centering
 \begin{tabular}[h]{|l|c|c|c|c|c|c|c|c||r}
  \hline
    Number of SRCs  & 2& 3& 4& 5& 6& 7& 8\\ \hline
Number of iterations &150&	300 &450 &780& 800 &810& 830\\ 
\hline
   \leavevmode\
  \end{tabular}

  \caption{Number of iterations for $A_{dist}$ convergence}
  \label{tab:1}
\end{table}

\subsubsection{Scenario 2 with 4 SRCs}
 \label{sec:sce-1-5}
 \leavevmode\

In the following, we will check if the fact of having more than two possible strategies per SRC could effect the results. To do so, we will consider $4$ SRCs defined as follows: $\alpha_1=0.9$, $\alpha_2=0.7$, $\alpha_3=1$ and $\alpha_4=0.9$.

\begin{figure}[Ht]
\begin{center}
\includegraphics[height=5cm,width=10cm]{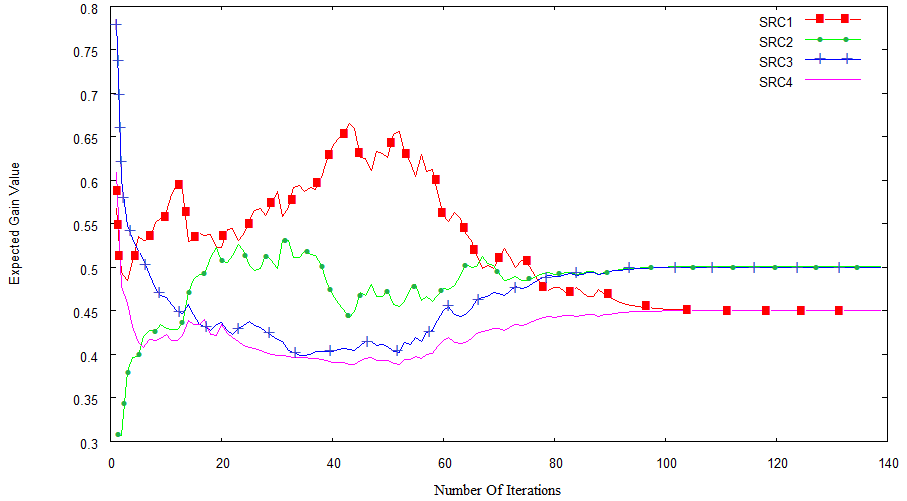}

\end{center}
\caption{Variation of SRCs expected gain for Scenario $2$}
\label{fig:7}
\end{figure}
 
\begin{figure}[Ht]
\begin{center}
\includegraphics[height=5cm,width=10cm]{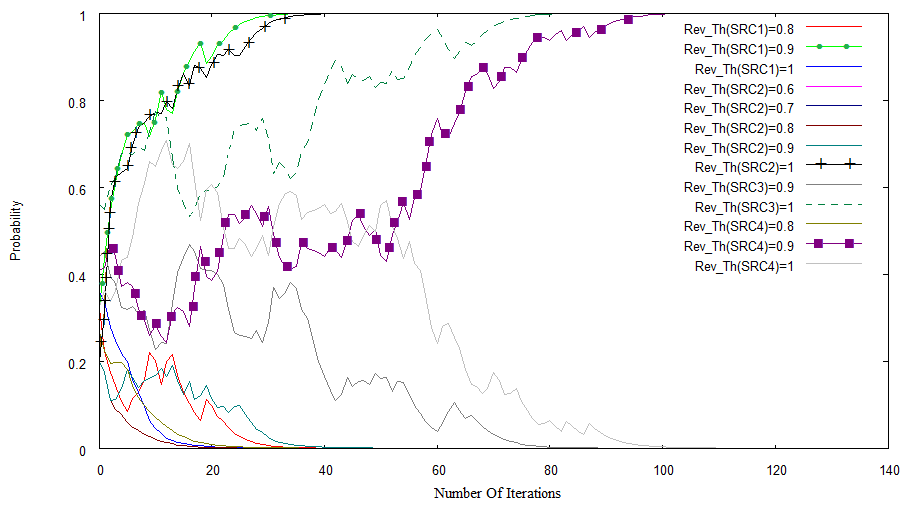}
\end{center}
\caption{Variation of SRCs strategies probabilities for Scenario $2$}
\label{fig:4}
\end{figure}

  $6$ pure Nash equilibriums are found analytically in the SRCs utility matrix and are the following: $$\Pi^*=\{(1,1,0.9,1,0.9); (0.9,0.9,1,1); (0.9,1,0.9,1);
(0.9,1,1,0.9); (1,0.9,0.9,1) ; (1,0.9,1,0.9); (1,1,0.9,0.9)\}.$$ 

For $A_{dist}$ algorithm, we will consider $b=0.1$ and $700$ iterations. In Figures \ref{fig:7} and \ref{fig:4}, each point represents the mean over 15 iterations of respectively SRCs expected gain value and SRCs strategy probability.

$A_{dist}$ converges after $500$ iterations. We notice in Figure \ref{fig:7} that the SRCs expected Gain stabilize as follows:

Expected gain(SRC$_1$)=Expected gain(SRC$_2$)=Expected gain(SRC$_4$)=0.45 and Expected gain(SRC$_3$)=Expected gain(SRC$_5$)=0.33

Even with more than two strategies per SRC, $A_{dist}$ still converges to a Pure Nash equilibrium. In this simulation, $\Pi^*=(0.9,1,1,0.9)$ is a pure Nash equilibrium. Figure  \ref{fig:4} shows that $SRC_{1}$, $SRC_{2}$, $SRC_{3}$ and $SRC_{4}$ needs respectively $175$, $200$, $410$ and $500$ iterations to learn the strategy maximizing its gain. 

When all the SRCs have pure strategies, we reach a stable state for the game restricted to SRCs. In order to make the system representing this game converge more rapidly, we will propose two solutions:
In the first one, we consider that the system is stable if all the SRCs have a strategy with a probability equal to p ($0<p<1$).

If we fix $p=0.8$, the software set up in the user equipment of $SRC_{1}$, $SRC_{2}$, $SRC_{3}$ and $SRC_{4}$ needs respectively 90, 90, 200 and 340 initiated connections to learn the strategy maximizing its gain. After 340 iterations, the system is considered as stable. Thus, the number of iterations for $A_{dist}$ convergence is reduced by $32\%$.

The second solution does not reduce the number of iterations for $A_{dist}$ convergence, but proposes to reduce its time duration (could also be seen as the number of requested connections). This solution is based on the fact of triggering the learning process several times per connection requested. This means that for a connection of duration D, the SRC's software will choose a new strategy following the updated strategy probability vector each d=q*D slot time ($q \in [0,1]$). Thus, the time duration (the number of requested connections) necessary for $A_{dist}$ convergence is reduced by q. 

We have focused on the convergence of $A_{dist}$ taking into consideration several strategies per SRC and $2 \leq N \leq 8$. We have found that $A_{dist}$ always converges to one among the pure Nash equilibriums detected analytically in the SRCs payoff matrix. To conclude, $A_{dist}$ permits to learn the game restricted to SRCs' pure Nash equilibrium whatever the SRCs profiles are.




\subsection{Simulation results}
\label{sec:conclusion Simulations}

Simulations results have shown that the game restricted to SRCs using a distributed learning algorithm converges to a pure Nash equilibrium. We have checked that this result is available for a number of SRCs varying from 2 to 8 for SRCs with exactly 2 strategies or more than 2 strategies. The distributed algorithm converges in a finite number of iterations. Two solutions has been proposed to reduce the number of requested connections necessary for this algorithm convergence. The distributed learning algorithm running in SRCs user equipments gives at convergence the minimum and the maximum amount of bandwidth to be requested by each SRC in Green and Yellow.

\section{Conclusion and perspectives}
\label{sec:perspectives}

In this article, we investigate the problem of sharing femto access taking a file transfer application as example. Only the competition between SRCs is modeled as a game considering a fixed SPC's bandwidth split. Our simulations focus on examples where SRCs objectives conflict: SRCs belonging to the same category of QoS sensitive SRCs.

Simulations have proved the efficiency of the Distributed Learning Algorithm: even if each SRC player has only local information, the algorithm always converges to a pure Nash equilibrium.

As a perspective, we will consider the sharing femto access game with several SRCs and several SPCs. The sharing femto access problem will thus be divided into two levels: a first level representing a game restricted to SRCs and a second level representing a game restricted to SPCs. We will study the properties of the second level game to check whether if they match with the game restricted to SRCs' properties.
We will also focus on the convergence time optimization of the algorithm $A_{Dist}$ applied on both SRCs game and SPCs game.


\begin{thebibliography}{1}

 
 
\bibitem{citation:6}
Sahand Haji~Ali Ahmad, Mingyan Liu, and Yunnan Wu.
\newblock Congestion games with resource reuse and applications in spectrum
  sharing.
\newblock {\em CoRR}, abs/0910.4214, 2009.

\bibitem{papa}
Niyaton Dusit and Ekram Hossain.
\newblock {\em Microeconomic Models for Dynamic Spectrum Management in
  Cognitive Radio Networks}, chapter~14, pages 391--423.
\newblock Springer, 2007.

\bibitem{citation:1}
David Hausheer, Nicolas~C. Liebau, Andreas Mauthe, Ralf Steinmetz, and Burkhard
  Stiller.
\newblock Token-based accounting and distributed pricing to introduce market
  mechanisms in a peer-to-peer file sharing scenario.
\newblock In {\em Proceedings of the 3rd International Conference on
  Peer-to-Peer Computing}, P2P '03, pages 200--, Washington, DC, USA, 2003.
  IEEE Computer Society.

\bibitem{RR:Mariem}
Mariem Krichen, Johanne Cohen, and Dominique Barth.
\newblock File transfer application for sharing femto access : Game properties.
\newblock Technical report, Universit\'e de Versailles, 2011.



\bibitem{citation:2}
John~F. Nash.
\newblock Equilibrium points in $n$-person games.
\newblock {\em Proceedings of the National Academy of Sciences of the United
  States of America}, 36:48--49, 1950.

\bibitem{AlgoGameTheory}
Nissan Nisan, Tim Roughgarden, {\'E}va Tardos, and Vijay~V. Vazirani.
\newblock {\em {Algorithmic Game Theory}}.
\newblock Cambridge University Press, 2007.

\bibitem{citation:3}
M.A.L.~Thathachar P.S.~Sastry, V.V.~Phansalkar.
\newblock Decentralized learning of {N}ash equilibria in multi-person
  stochastic games with incomplete information.
\newblock {\em IEEE transactions on system, man, and cybernetics}, 24(5), 1994.

\bibitem{rosenthal73}
R.W. Rosenthal.
\newblock {A class of games possessing pure-strategy Nash equilibria}.
\newblock {\em International Journal of Game Theory}, 2(1):65--67, 1973.

\bibitem{citation:5}
Takashi Ui.
\newblock A shapley value representation of potential games.
\newblock {\em Games and Economic Behavior}, 31(1):121--135, April 2000.

\bibitem{citation:7}
Yiping Xing, Chetan~N .Mathur, M.~A Haleem, R.~Chandramouli, and K.~P
  Subbalakshmi.
\newblock Real-time secondary spectrum sharing with qos provisioning.
\newblock In {\em Consumer Communications and Networking Conference (CCNC)},
  pages 630 -- 634, 2006.


\end{thebibliography}
\end{document}